# Key Exchange Protocol in the Trusted Data Servers Context

**Quoc-Cuong To, Benjamin Nguyen, Philippe Pucheral**

PRiSM Laboratory, Univ. of Versailles, France
<Fname.Lname>@prism.uvsq.fr

INRIA Rocquencourt, Le Chesnay, France
<Fname.Lname>@inria.fr

# Contents

# 1   Introduction

With the convergence of mobile communications, sensors and online social networks technologies, an exponential amount of personal data - either freely disclosed by users or transparently acquired by sensors - end up in servers. This massive amount of data, the *new oil,* represents an unprecedented potential for applications and business (e.g., car insurance billing, traffic decongestion, smart grids optimization, healthcare surveillance, participatory sensing). However, centralizing and processing all one's data in a single server incurs a major problem with regards to privacy. Indeed, individuals' data is carefully scrutinized by governmental agencies and companies in charge of processing it [de Montjoye *et al.* 2012]. Privacy violations also arise from negligence and attacks and no current server-based approach, including cryptography based and server-side secure hardware [Agrawal *et al.* 2002], seems capable of closing the gap. Conversely, decentralized architectures (e.g., personal data vault), providing better control to the user over the management of her personal data, impede global computations by construction.

To reconcile privacy protection and global computation to the best benefit of the individuals, the community and the companies, [To *et al.* 2014] capitalizes on a novel architectural approach called *Trusted Cells* [Anciaux *et al.* 2013]. Trusted Cells push the security to the edges of the network, through personal data servers [Allard *et al.* 2010] running on secure smart phones, set-top boxes, plug computers[1] or secure portable tokens[2] forming a global decentralized data platform. Indeed, thanks to the emergence of low-cost secure hardware and firmware technologies like ARM TrustZone[3], a full Trusted Execution Environment (TEE) will soon be present in any client device.

As discussed in [Anciaux *et al.* 2013], trusted hardware is more and more versatile and has become a key enabler for all applications where trust is required at the edges of the network. Figure 1 depicts different scenarios where a Trusted Data Server (TDS) is called to play a central role, by reestablishing the capacity to perform global computations without revealing any sensitive information to central servers. TDS can be integrated in energy smart meters to gather energy consumption raw data, to locally perform aggregate queries for billing or smart grid optimization purpose and externalize only certified results, thereby reconciling individuals' privacy and energy providers' benefits. Green button[4] is another application example where individuals accept sharing their data with their neighborhood through distributed queries for their personal benefit. Similarly, TDS can be integrated in GPS trackers to protect individuals' privacy while securely computing insurance fees or carbon tax and participating in general interest distributed services such as traffic jam reduction. Moreover, TDSs can be hosted in personal devices to implement secure personal folders like e.g., PCEHR (Personally Controlled Electronic Health Record) fed by the individuals themselves thanks to the Blue Button initiative[5] and/or quantified-self devices. Distributed queries are useful in this context to help epidemiologists performing global surveys or allow patients suffering from the same illness to share their data in a controlled manner.

---
[1] http://freedomboxfoundation.org/
[2] http://www.gd-sfs.com/portable-security-token
[3] http://www.arm.com/products/processors/technologies/trustzone.php

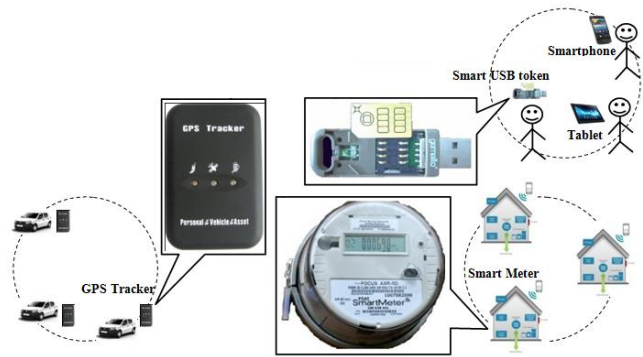

Fig. 1. Different scenarios of TDSs

The architecture [To *et al.* 2014] considers is decentralized by nature. It is formed by a large set of low power TDSs embedded in secure devices. Despite the diversity of existing hardware platforms, a secure device can be abstracted by (1) a Trusted Execution Environment and (2) a (potentially untrusted but cryptographically protected) mass storage area (see Fig. 2)[6]. E.g., the former can be provided by a tamper-resistant microcontroller while the latter can be provided by Flash memory. The important assumption is that the TDS code is executed by the secure device hosting it and thus cannot be tampered, even by the TDS holder herself. Each TDS exhibits the following properties: High security, modest computing resource, and low availability.

---

[4] http://www.greenbuttondata.org/
[5] http://healthit.gov/patients-families/your-health-data
[6] For illustration purpose, the secure device considered in our experiments is made of a tamper-resistant microcontroller connected to a Flash memory chip.

Since TDSs have limited storage and computing resources and they are not necessarily always connected, an external infrastructure, called hereafter *Supporting Server Infrastructure* (SSI), is required to manage the communications between TDSs, run the distributed query protocol and store the intermediate results produced by this protocol. Because SSI is implemented on regular server(s), e.g., in the Cloud, it exhibits the same low level of trustworthiness, high computing resources, and availability.

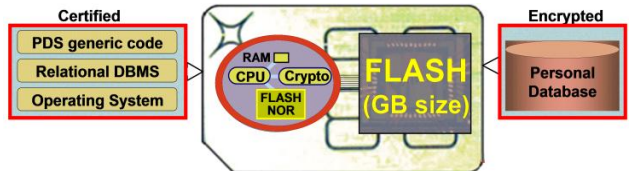

Fig. 2. Trusted Data Servers

Hence, even if there exist differences among Secure Devices (e.g., smart tokens are more robust against tampering but less powerful than TrustZone devices), all provide *much stronger security guarantees* combined with a *much weaker availability and computing power* than any traditional server.

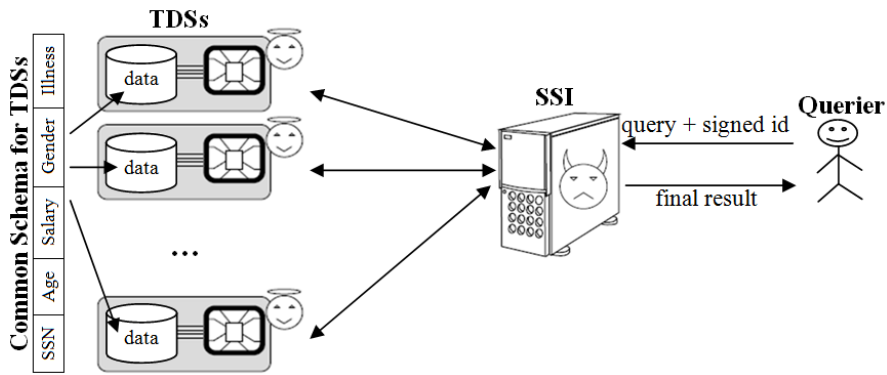

Fig. 3. The Asymmetric Architecture

The computing architecture, illustrated in Fig. 3 is said *asymmetric* in the sense that it is composed of a very large number of low power, weakly connected but highly secure TDSs and of a powerful, highly available but untrusted SSI.

TDSs are the unique elements of trust in the architecture and are considered *honest*. As mentioned earlier, no trust assumption needs to be made on the TDS holder herself because a TDS is tamper-resistant and enforces the access control rules associated to its holder (just like a car driver cannot tamper the GPS tracker installed in her car by its insurance company or a customer cannot gain access to any secret data stored in her banking smartcard).

We primarily consider *honest-but-curious* (also called *semi-honest*) SSI (i.e., which tries to infer any information it can but strictly follows the protocol), concentrating on the prevention of confidentiality attacks.

The objective of [To *et al.* 2014] is to implement a querying protocol so that (1) the querier can gain access only to the final result of authorized queries (not to the raw data participating in the computation), as in a traditional database system and (2) intermediate results stored in SSI are obfuscated. To ensure the confidentiality of this protocol, the shared keys between Querier and TDSs play an important role, and therefore an adaptive key exchange protocol suitable to TDSs context is necessary.

The aim of this technical report is to complement the work in [To *et al.* 2014] by proposing a Group Key Exchange protocol so that the Querier and TDSs (and TDSs themselves) can securely create and exchange the shared key. Then, the security of this protocol is formally proved using the game-based model. Finally, we perform the comparison between this protocol and other related works.

## 2   State-of-the-Art on Group Key Management

Group Key Exchange (GKE) protocols can be roughly classified into three classes: centralized, decentralized, and distributed [Rafaeli and Hutchison 2003]. In centralized group key protocols, a single entity is employed to control the whole group and is responsible for distributing group keys to group members. In the decentralized approaches, a set of group managers is responsible for managing the group as opposed to a single entity. In the distributed method, group members themselves contribute to the formation of group keys and are equally responsible for the re-keying and distribution of group keys. Their analysis [Rafaeli and Hutchison 2003] made clear that there is no unique solution that can satisfy all requirements. While centralized key management schemes are easy to implement, they tend to impose an overhead on a single entity. Decentralized protocols are relatively harder to implement and raise other issues, such as interfering with the data path or imposing security hazards on the group. Finally, distributed key management, by design, is simply not scalable. Hence it is important to understand fully the requirements of the application to select the most suitable GKE protocol. Under the computational Diffie-Hellman assumption, some works [Wu *et al.* 2011, Bresson *et al.* 2004] proposed group key exchange protocol suitable for low-power devices. These works achieve communication efficiency because they require only two communication rounds to establish the shared key. They also require little computing resources of participants and are thus suitable for the Trusted Data Server (TDS) context [To *et al.* 2014b].

# 3 Overview of Key Management

There are numerous ways to share the keys between TDSs and Querier depending on which context we consider.

In the closed context, we assume that all TDSs are produced by the same provider, so the shared key $k_T$ can be installed into TDSs at manufacturing time. If Querier also owns a TDS, key $k_Q$ can be installed at manufacturing time as well. Otherwise, Querier must create a private/public key and can use another way (PKI or GKE described below) to exchange key $k_Q$. An illustrative scenario for the closed context can be: patients and physicians in a hospital get each a TDS from the hospital, all TDSs being produced by the same manufacturer, so that the required cryptographic material is preinstalled in all TDSs before queries are executed.

In an open context, a Public Key Infrastructure (PKI) can be used so that queriers and TDSs all have a public-private key pair. When a TDS or querier registers for an application, it gets the required symmetric keys encrypted with its own public key. Since the total number of TDS manufacturers is assumed to be very small (in comparison with the total number of TDSs) and all the TDSs produced by the same producer have the same private/public key pair, the total number of private/public key pairs in the whole system is not big. Therefore, deploying a PKI in our architecture is suitable since it does not require an enormous investment in managing a very large number of private/public key pairs (i.e., proportional to the number of TDSs). PKI can be used to exchange both keys $k_Q$ and $k_T$ for both Querier cases i.e. owning a token or not. In the case we want to exchange $k_T$, we can apply the above protocol for $k_Q$ with Querier being replaced by one of the TDSs. This TDS can be chosen randomly or based on its connection time (e.g., the TDS that has the longest connection time to SSI will be chosen).

An illustrative scenario for the open context can be: TDSs are integrated in smart phones produced by different smart phone producers. Each producer has many models (e.g., iPhone 1-6 of Apple, Galaxy S1-S5 of Samsung, Xperia Z1-Z4 of Sony…) and we assume that it installs the same private/public key on each model. In total, there are about one hundred models in the current market, so the number of different private/public keys is manageable. The phone's owner can then securely take part in surveys such as: what is the volume of 4G data people living in Paris consume in one month, group by network operators (Orange, SFR…).

In PKI, only one entity creates the whole secret key, and securely transfers it to the others. In the distributed key agreement protocols, however, there is no centralized key server available. This arrangement is justified in many situations—e.g., in peer-to-peer or ad hoc networks where centralized resources are not readily available or are not fully trusted to generate the shared key entirely. Moreover, an advantage of distributed protocols over the centralized protocols is the increase in system reliability, because the group key is generated in a shared and contributory fashion and there is no single-point-of-failure [Lee *et al*. 2006]. Group AKE protocols are essential for secure collaborative (peer-to-peer) applications [Lee *et al*. 2006]. In these circumstances, every participant wishes to contribute part of its secrecy to generate the shared key such that no party can predetermine the resulting value. In other words, no party is allowed to choose the group key on behalf of the whole group. These reasons lead to another way to exchange the shared key between TDSs and Querier in the open context. In this way, we use the GKE [Wu *et al.* 2011, Amir *et al.* 2004, Wu *et al.* 2008] so that Querier can securely exchange the secret contributive key to all TDSs. Some GKE protocols [Amir *et al.* 2004] require a broadcast operation in which a participant sends part of the key to the rest. These protocols are not suitable for our architecture since TDSs communicate together indirectly through SSI. This incurs a lot of operations for SSI to broadcast the messages (i.e., $O(n^2)$, with *n* is the number of participants). Other protocols [Wu *et al.* 2008] overcome this weakness by requiring that participants form a tree structure to reduce the communication cost. Unfortunately, SSI has no knowledge in advance about TDSs thus this tree cannot be built. The work in [Wu *et al.* 2011] proposes a protocol with two rounds of communications and only one broadcast operation. However, this protocol still has the inherent weakness of the GKE: all participants must connect during the key exchange phase. This characteristic does not fit in our architecture since TDSs are weakly connected. Finally, the Broadcast Encryption Scheme (BES) [Castelluccia *et al.* 2005] requires that all participants have a shared secret in advance, preventing us from using it in a context where TDSs are produced by different manufacturers.

In consequence, we propose an adaptive GKE scheme, fitting our architecture as follows.

## 4 The Adaptive Key Exchange Protocol

Let p, q be two large primes satisfying $p = 2q + 1$; $G_q$ be a subgroup of $Z_p^*$ with the order q; g be a generator of the group $G_q$; $H_1$, $H_2$ be two one-way hash functions such that $H_1, H_2: \{0, 1\}^* \rightarrow Z_q^*$; SID be a public session identity (note that each session is assigned a unique SID). Without loss of generality, let $\{Q, U_1, U_2,..., U_n\}$ be a set of participants who want to generate a group secret key, where **Q** is the Querier and $U_1, U_2,..., U_n$ are TDSs. This dynamic GKE protocol is depicted in Fig. 7 and the detailed steps are described as follows:

*Step 1*: Each client $U_i$ ($1 \leq i \leq n$) computes $r_i = MAC(SID, Kp_i)$ which is the Message Authentication Code (MAC) of the session identity SID, using a key derived from $Kp_i$ and $z_i = g^{r_i} \mod p$, as well as a signature $\sigma_i$ of $z_i$ under $Kp_i$ which is the private key of $TDS_i$. Then, each $U_i$ sends ($\sigma_i$, $z_i$) to SSI. Since all TDSs produced by the same producer share the same private/public key pair ($Kp_i$, $Kpu_i$), they generate the same $z_i$. When this collection phase stops, SSI forwards all these ($\sigma_i$, $z_i$) to Querier Q.

*Step 2*: Querier Q first selects two random values $r_0, r \in Z_q^*$ and computes $z_0 = g^{r_0} \mod p$. Upon receiving n pairs ($\sigma_i$, $z_i$) ($1 \leq i \leq n$), Querier eliminates the duplicated $z_i$, (we assume that there remains only m pairs ($\sigma_i$, $z_i$) with distinct $z_i$). Since the number of producers is very small in comparison with the number of TDSs, we have m << n. For each m pairs ($\sigma_i$, $z_i$), Querier Q checks the signature $\sigma_i$ using public key

Kpu$_i$ of each U$_i$ and if they are all correct, Q computes x$_i$ = z$_i^{r0}$ mod p and y$_i$ = H$_2$(x$_i$ || SID) ⊕ r for i=1, 2,..., m. Finally, Q computes the shared session key SK = H$_2$(r||y$_1$||y$_2$||...||y$_m$||SID), a signature o$_0$ of z$_0$ and broadcasts (o$_0$, y$_1$, y$_2$..., y$_m$, z$_0$, SID) to all TDSs. Since m << n, the length of the broadcast message (o$_0$,y$_1$,y$_2$...,y$_m$,z$_0$, SID) is very short, saving network bandwidth.

*Step 3*: Upon receiving the messages (o$_0$, y$_1$, y$_2$..., y$_m$, z$_0$, SID), each TDS$_i$ verifies the signature o$_0$ and then can compute x$_i$ and r = y$_i$ ⊕ H$_2$(x$_i$ || SID) and uses r to obtain the shared key SK = H$_2$(r||y$_1$||y$_2$||...||y$_m$||SID). In this step, even if some TDSs did not participate in the first step of the protocol, they still can get the secret group key SK because they can use their private key and the public hash function H$_1$ to compute the value r$_i$ that all the TDSs belonging to the same manufacturer can compute. This characteristic reflects the adaptive property of this protocol since it is suitable for our context in which all participants do not need to connect during the key exchange phase.

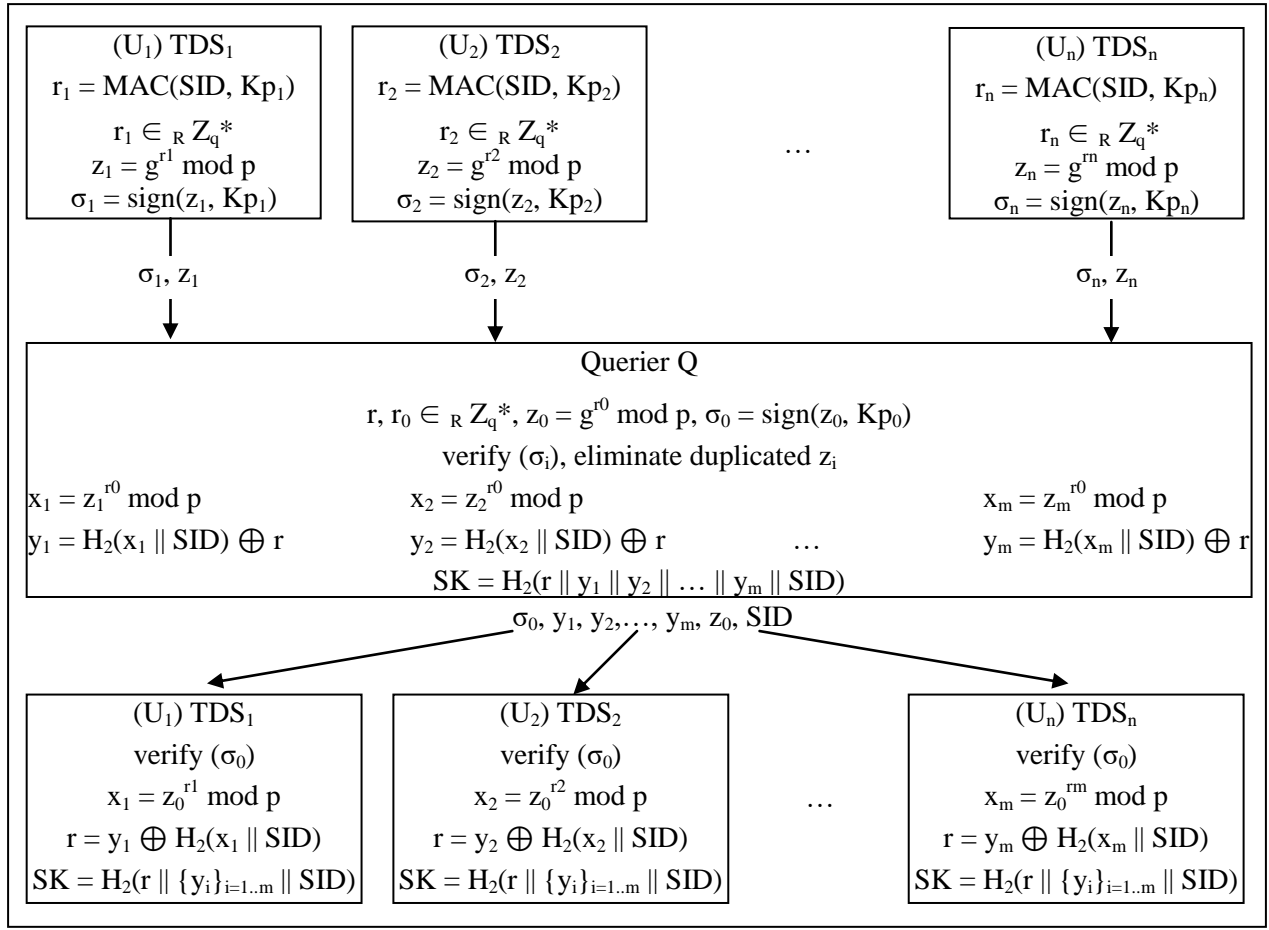

Fig. 7. Adaptive Group Key Exchange Protocol

# 5   Security Analysis of the Adaptive Key Exchange Protocol

Formally, there are five notions of security [Bresson and Manulis 2007] used as the standards to evaluate the security of a GKE protocol as follows:
- Authenticated key exchange (AKE): requires the indistinguishability of computed group keys from random keys.
- Forward/backward secrecy [Wu *et al.* 2011]: requires that other participants need not to re-run the protocol when a participant joins/leaves the group.
- Contributiveness: requires that all protocol participants equally contribute to the computation of the group key. These requirements implicitly state a difference between PKI and GKE protocols in the spirit of generating the key
- Universal Composability (UC) [Katz and Shin 2005]: requires that protocol is secure against insider impersonation attacks.
- Mutual Authentication (MA): requires that all protocol participants are ensured to actually compute the same key.

Since our protocol is closely adapted from protocol in [Bresson *et al.* 2004] which is already formally proved to be AKE-secure, we just point out the difference between these two protocols and show that this difference does not affect the security of protocol. To do that, we first prove that the secret value $r_i$ generated by each TDS is indistinguishable from random number, then we use the game-based model inspired by [Bresson *et al.* 2004] to formally prove that our protocol is AKE-secure.

The only difference between the two protocols is the way the secret value $r_i$ (that contributes to the shared key) in each TDS is generated. In Bresson's protocol, the value $r_i$ is randomly generated. In our solution, TDSs cannot simply randomly generate $r_i$ because the adaptive property (i.e., all TDSs from a same manufacturer generate the same $r_i$ so that not all participants need to connect during the key exchange phase) will be lost. The solution we propose is to compute $r_i$ as the MAC of an SID, using a key derived from $Kp_i$: $r_i$ = MAC(SID, $Kp_i$). This way, all TDSs from the same manufacturer will return the same $r_i$ for a given SID because they possess the same $Kp_i$, and as long as a querier cannot replay an SID[7], the answers will be unpredictable for the querier. Since SID is random for each session and the value $r_i$ is derived from secret key $Kp_i$ and the SID using a random oracle, MAC(SID, $Kp_i$) is indistinguishable from random $r_i$. In other words, our scheme is indistinguishable from the original scheme of Bresson.

The philosophy of this proof is largely inspired by [Bresson *et al.* 2004] and part of the text is borrowed from this paper. For the detail formal proof of security, interested readers can refer to [Bresson *et al.* 2004].

Note that for the security of the proposed protocol, given the Diffie-Hellman problem (see below), we make the following classical DDH and CDH assumptions, and assume there exists a secure one-way hash function.

---

[7] We need to put in place a mechanism to ensure that SIDs cannot be reused. For example, TDSs could store them in their internal DBs and only reply to queries using a fresh SID.

**Decision Diffie-Hellman (DDH) problem**: Given $y_a = g^{x1}$ mod p and $y_b = g^{x2}$ mod p for some $x1, x2 \in Z^*_q$, the DDH problem is to distinguish two tuples ($y_a$, $y_b$, $g^{x1x2}$ mod p) and ($y_a$, $y_b$, $R \in \mathbb{G}_q$) where R is a random value in a (multiplicative) cyclic group $\mathbb{G}_q$ of order q.

**DDH assumption**: There exists no probabilistic polynomial-time algorithm can solve the DDH problem with a non-negligible advantage.

**Computational Diffie-Hellman (CDH) problem**: Given a tuple (g, $g^{x1}$ mod p, $g^{x2}$ mod p) for some $x1, x2 \in Z^*_q$, the CDH problem is to compute the value $g^{x1x2}$ mod $p \in G_q$.

**CDH assumption**: There exists no probabilistic polynomial-time algorithm can solve the CDH problem with a non-negligible advantage. Formally, a (t, $\varepsilon$)-CDH attacker in $\mathbb{G}$ is a probabilistic machine $\Delta$ running in time t such that:

$Succ_\mathbb{G}^{cdh}(\Delta) = Pr_{x1,x2}[\Delta(g^{x1}, g^{x2}) = g^{x1x2}] \geq \varepsilon$

The CDH problem is (t, $\varepsilon$)-intractable if there is no (t, $\varepsilon$)-attacker in $\mathbb{G}$. The CDH assumption states that is the case for all polynomial t and any non-negligible $\varepsilon$.

**Hash function assumption**: A secure one-way hash function H: X={0,1}* -> Y=$Z^*_q$ must satisfy following requirements:

(i) for any $y \in Y$, it is hard to find $x \in X$ such that H(x)=y.
(ii) for any $x \in X$, it is hard to find $x' \in X$ such that $x' \neq x$ and H(x') = H(x).
(iii) it is hard to find x, $x' \in X$ such that $x' \neq x$ and H(x)=H(x').

**Adversarial Model**: we consider an adversary $\mathbb{A}$ which is a Probabilistic Polynomial-Time (PPT) algorithm having complete control over the network. $\mathbb{A}$ can invoke protocol execution and interact with protocol participants via queries to their oracles. We model the capabilities of A through the following queries:

- Adversary $\mathbb{A}$ can send arbitrary messages to Q using the SendQ-query.
- Adversary $\mathbb{A}$ can send arbitrary messages to $TDS_i$ using the Send-query.
- Known-key attacks are modeled by the Reveal-query. This query allows the adversary $\mathbb{A}$ to learn the value of a particular session key SK if participant has computed the key.
- Test-query: is used to model the AKE-security of a protocol. This query is answered as follows: The oracle generates a random bit b. If b = 1 then $\mathbb{A}$ is given session key SK, and if b = 0 then $\mathbb{A}$ is given a random string.

To prove that the proposed protocol is AKE-secure, we apply the Bresson's Game-based security model [Bresson *et al*. 2004] by using a sequence of games $G_0$ through $G_3$, in which we simulate the protocol and consider $\mathbb{A}$ attacking the simulated protocol.

We denote by b the bit involved in the Test-query, by b' the guess output by $\mathbb{A}$, by $q_s$ the total number of Send-queries asked to the participants. We refer in game $G_i$ the event $S_i$ as being b=b'. When Test-query is asked, $\mathbb{A}$ gets back either SK if bit b=1 or a random string of same length if b=0. When $\mathbb{A}$ terminates, it outputs a single bit b'. Semantic security means that $\mathbb{A}$ does not learn any information about SK and has no advantage to guess the bit b. So we define the advantage of adversary $\mathbb{A}$ to guess bit b for our protocol P in the game as:

$$\text{Adv}_P^{ake}(A) = |\text{Pr}_{b'}[b'=1 | | b=1] - \text{Pr}_{b'}[b'=1 | | b=0]| = 2\,\text{Pr}_{b,b'}[b=b'] - 1.$$

A protocol P is an $(t, \varepsilon)$-secure AKE if $\text{Adv}_P^{ake}(A)$ is less than $\varepsilon$ for all probabilistic adversary $A$ which running time is bounded by $t$.

**Game $G_0$.** The Querier Q is given $z_0 = g^{r_0} \bmod p$ and each TDS is given a pair of private/public key, and computes $z_i$. We thus have:

$$\text{Pr}[S_0] = (\text{Adv}_P^{ake}(A) + 1)/2 \qquad (1)$$

**Game $G_1$.** The security notion for a signature scheme is that it is computationally infeasible for an adversary to produce a valid forgery $\sigma$ with respect to any message $m_e$ under a chosen-message attack (CMA) [Bresson and Manulis 2007]. It is $(t, q_s, \varepsilon)$–CMA-secure if there is no adversary $A$ which can get a probability greater than $\varepsilon$ (denoted as $\text{Succ}_{SIGN}^{cma}(A)$) in mounting an existential forgery under a CMA attack within time $t$, after $q_s$ signing queries. We refer to Forge as the event that $A$ asks for a SendQ($m_e'$)-query, such that the verification of the signature is correct and $m_e'$ was not previously output by a client as an answer to another Send-query. In this case, we abort the game and fix b' randomly. The game $G_1$ and $G_0$ are similar as long as Forge does not happen. By guessing the impersonated client, one easily gets:

$$|\text{Pr}[S_1] - \text{Pr}[S_0]| \leq \text{Pr}[\text{Forge}] \leq m_e * \text{Succ}_{SIGN}^{cma}(t, q_s) \qquad (2)$$

**Game $G_2$.** We are given a Diffie-Hellman triple ($A = g^\alpha$, $B = g^\beta$, $C = g^\gamma$) with the values $\alpha$, $\beta$ (and thus $\gamma = \alpha * \beta \bmod q$), and define $r \leftarrow \alpha$, $z \leftarrow A = g^\alpha$. Furthermore, exponent $r_i$ is defined by $\beta + \sigma_i \bmod q$, and $z_i \leftarrow B g^{\sigma_i}$. So, $\alpha_i$ is set to $C A^{\sigma_i}$:

$$\text{Pr}[S_2] = \text{Pr}[S_1] \qquad (3)$$

**Game $G_3$.** Any hash value involving an $x_i$ or $y_i$ (either $H_2(x_i | | SID)$ or $H_2(r | | \{y_i\}_{i=1..m} | | SID)$) asked by TDSs or Q are answered independently from the random oracles. Since the same hash queries, asked by the adversary, are still answered by querying the random oracles, some inconsistency may occur. Such an inconsistency is discovered by the adversary if such a hash query is asked by the adversary, event which we denote by AskH:

$$|\text{Pr}[S_3] - \text{Pr}[S_2]| \leq \text{Pr}[\text{AskH}] \qquad (4)$$

Such an event AskH means that some $x_i$ (among at most $q_s$) appears in the list of the hash queries. By guessing the $\alpha_i$ instance that has been asked by the adversary, and the corresponding hash query, one can extracts $C = \alpha_i * A^{-\sigma_i}$:

$$\text{Pr}[\text{AskH}] \leq q_H q_s * \text{Succ}_6^{cdh}(t) \qquad (5)$$

with $q_s$ active requests, and $q_H$ queries to the hash oracles.

From (1), (2), (3), (4) and (5), we conclude that:

$$\text{Adv}_P^{ake}(A) \leq 2 m_e * \text{Succ}_{SIGN}^{cma}(t, q_s) + q_H q_s * \text{Succ}_6^{cdh}(t).$$

This inequality shows that our protocol P is AKE secure, according to the AKE definition above.

To prove the contributiveness, we can see that after Q broadcasts ($\sigma_0$, $y_1$, $y_2$..., $y_m$, $z_0$, SID) to all TDSs, each TDS can use its own secret $r_i$ to compute the value $r$ and then obtains an identical group key SK. This means the following equations hold:

$$SK = H_2(r | | \{y_i\}_{i=1..m} | | SID) = H_2(y_1 \oplus H_2(z_0^{r_1} | | SID) | | \{y_i\}_{i=1..m} | | SID)$$
$$= H_2(y_2 \oplus H_2(z_0^{r_2} | | SID) | | \{y_i\}_{i=1..m} | | SID)$$
$$= H_2(y_m \oplus H_2(z_0^{r_m} | | SID) | | \{y_i\}_{i=1..m} | | SID)$$

Set $V = y_1 \oplus H_2(z_0^{r_1} || SID) = y_2 \oplus H_2(z_0^{r_2} || SID) = ... = y_m \oplus H_2(z_0^{r_m} || SID)$. It implies $y_1 = H_2(z_0^{r_1} || SID) \oplus V$, $y_2 = H_2(z_0^{r_2} || SID) \oplus V$, ..., $y_m = H_2(z_0^{r_m} || SID) \oplus V$. Obviously, each $y_i$ includes the participant TDS's secret value $r_i$ for i=1..m. By the group key $SK = H_2(r || \{y_i\}_{i=1..m} || SID) = H_2(y_1 \oplus H_2(z_0^{r_1} || SID) || \{y_i\}_{i=1..m} || SID)$, each participant ensures that his contribution has been involved in the group key SK, providing contributiveness of the protocol.

To prove the forward/backward secrecy for member joining/leaving, we first adopt this lemma (proved in [Wu *et al.* 2011]): Assume that three secret parameters a, b, and c are randomly selected from $Z_p^*$. If an adversary knows two values $H(a) \oplus b$ and $H(a) \oplus c$, then the secret a and b are not computable under the hash function assumption.

Assume a new TDS $U_{m+1}$ belonging to a new manufacturer wants to join the group. $U_{m+1}$ sends $z_{m+1} = g^{r_{m+1}} \mod p$ to Q. Then, Q selects $r' \in Z_q^*$ and computes $y'_i$ (i=1..m+1) with a new SID'. Finally, Q broadcasts ($\sigma_0$, $y'_1$, $y'_2$..., $y'_{m+1}$, $z_0$, SID') to all TDSs. All TDSs can compute a new group key $SK' = H_2(r' || y'_1 || y'_2 || ... || y'_{m+1} || SID')$ with $r' = y'_i \oplus H_2(z_0^{r_i} || SID')$. We want to prove that the new TDS $U_{m+1}$ cannot compute the previous group key $SK = H_2(r || y_1 || y_2 || ... || y_m || SID)$. $U_{m+1}$ can record all the previous transmitted messages ($z_i = g^{r_i} \mod p$, $y_i$, SID) for i=0..m. Obviously, $U_{m+1}$ can compute the key SK only when he can get the value r or $x_i$.

In the first case, due to the CDH assumption, it is hard to compute $g^{r_0 * r_i} \mod p = x_i$, given a tuple (g, $z_0 = g^{r_0} \mod p$, $z_i = g^{r_i} \mod p$), making it impossible for $U_{m+1}$ to obtain $x_i$ from $z_i$.

In the second case, $U_{m+1}$ can get the value r or $x_i$ from ($y_i$, $y'_i$, SID, SID') for i=1..m, where $y_i = H_2(x_i || SID) \oplus r$ and $y'_i = H_2(x_i || SID') \oplus r'$. Without loss generality, we set $a = x_i || SID = x_i || SID'$, b=r, and c=r' such that $y_i = H(a) \oplus b$ and $y'_i = H(a) \oplus c$. Applying above lemma, we prove that the values a and b are not computable under the hash function assumption. Hence, obtaining the value r or $x_i$ is also impossible.

In both cases, $U_{m+1}$ cannot compute the previous group key SK, proving the forward secrecy of our protocol. By applying the same technique, we can prove that our protocol also provides backward secrecy.

In short, our protocol achieves two goals: (i) Each TDS has to generate random secret values (to ensure that the protocol is AKE-secure), and (ii) all TDSs from a same manufacturer contribute the same secret value (to guarantee the adaptivity of our protocol in the TDSs context).

While AKE-security became meanwhile standard to prove security of GKE, it does not take into account any notion of protection against insider attacks, and thus AKE-secure protocols may be completely insecure against attacks by malicious insiders [Katz and Shin 2005]. To encompass these attacks, [Katz and Shin 2005] proposed a compiler that can convert any AKE-secure GKE protocol into an UC-secure GKE protocol which ensures security against insider attacks.

This compiler was then extended by [Bresson and Manulis 2007] to support MA-security. This extended compiler, called `C-MACON`, can be used to transform any AKE-secure GKE protocol into a GKE protocol which is additionally MA-secure.

Seeing GKE protocols as building blocks for high-level applications, these two add-on compilers allow to enhance security of a protocol in a black-box manner, that is, independently of the implementation of the protocol being enhanced [Bresson et al. 2007]. In other words, a security-enhancing GKE compiler C is a procedure which takes as input a GKE protocol P and outputs a compiled GKE protocol $C_P$ with additional security properties possibly missing in P.

Since `C-MACON` is the extension of compiler in [Katz and Shin 2005], we can apply this compiler to our protocol P, which is already AKE-secure as proved above, to create a new GKE protocol `C-MACON`$_P$ that achieves all notions of security listed above. Then, the interesting question is how much overhead incurred for the compiled GKE protocol `C-MACON`$_P$ in compared with the original protocol P. `C-MACON`$_P$ achieves MA-security at an additional cost of only two communication rounds, which is an acceptable overhead. The following section details this overhead on each TDS when we apply this compiler.

Note that, even if SSI also possesses a TDS, it still cannot access the key shared between TDSs. As stated above, TDS code and content cannot be tampered, even by its holder. The only information that SSI in possession of a TDS can see is a stream of encrypted tuples [To *et al.* 2014b].

## 6 The Efficiency of the Adaptive Key Exchange Protocol

This method has two advantages in terms of asynchronous connection and performance over other GKEs in literature. First, this adaptive protocol perfectly fits our weakly connected assumption regarding the participating TDSs. Specifically, this protocol does not require that all TDSs connect at the same time to form the group, the connection of a single TDS per manufacturer being enough. The encrypted $k_T$ could be stored temporarily on SSI so that the offline TDS can get it as soon as it comes online and still take part in the protocol (i.e., any TDS that connects later can use its private key to compute the $r_i$, then *SK*, and after that can participate into the computation). Second, even if a TDS opts out of a SQL query in the collection phase, it can still contribute to the parallel computation in the aggregation phase. With a traditional distributed key exchange, any TDS disconnected during setup will require a new key exchange to take place. With our protocol, each TDS contributes to part of the shared secret key, the only requirement is that at least one TDS per manufacturer participates in step 1 to contribute to the value $r_i$ representing this manufacturer.

In terms of performance, this protocol is not a burden because it requires only 2-round of communications as shown in Fig. 7. Furthermore, the first round can be combined with the collection phase, helping reduce the protocol to only one phase.

To compare our protocol with other solutions in literature in term of performance, we first denote the following notations:

$T_{sig}$: time to execute a signature operation (signing or verifying).

$T_{exp}$: time to execute an exponentiation operation.

$T_H$: time to execute an one-way hash function.

$T_{per}$: time to execute an one-way permutation (used in `C-MACON`).

$T_{psf}$: time to execute a pseudorandom function (used in `C-MACON`).

We compare communication cost (i.e., number of rounds) and computational cost on each TDS among four protocols: our protocol (P), our protocol plus `C-MACON` compiler applied to our protocol (P + `C-MACON`$_P$), Bresson's protocol ($P_B$) in [Bresson et al. 2004], and Wu's protocol ($P_W$) in [Wu et al. 2011]. These protocols have one thing in common: they address the GKE protocols for low power mobile devices. Table 1 summarizes the comparison among these protocols.

|  | $P_B$ | $P_W$ | P | P + `C-MACON`$_P$ |
|---|---|---|---|---|
| Number of rounds | 2 | 2 | 2 | 4 |
| Computational cost on each TDS | $2T_{exp} + T_H + T_{sig}$ | $2T_{exp} + 2T_H$ | $2T_{exp} + 3T_H + 2T_{sig}$ | $2T_{exp} + 3T_H + (m+3)T_{sig} + mT_{per} + mT_{psf}$ |

Table 1. Performance comparison among GKE protocols.

From Table 1, it is easy to see that to obtain stronger notions of security, each participant in that protocol must spend more cost for hashing, verifying the signatures and other operations.

Similar to PKI, adaptive GKE can be used to exchange keys $k_Q$ and $k_T$ in both cases of Querier. However, although PKI and GKE are both based on the private/public keys in the open context, they differ in the way to generate the shared key. PKI is centralized and needs to trust the certification authority (which is a single point of attack) to generate the shared key. In contrast, with the adaptive GKE every TDS contributes part of the secret to generate the shared key.